\documentclass[10pt, conference, letterpaper]{IEEEtran}

\usepackage[markup=underlined]{changes}

\usepackage{url}              		 
\usepackage{amsfonts}       	
\usepackage{nicefrac}       	  

\usepackage{graphicx}
\usepackage{color}
\usepackage{amsmath,amsthm,amssymb, amsfonts}
\usepackage{algorithm, algorithmic}
\usepackage{bbm,dsfont}
\usepackage{enumerate, cases}
\usepackage{thmtools,thm-restate}

 \usepackage{multirow}

\usepackage{dblfloatfix}

\usepackage{hyperref}		 
\hypersetup{
	colorlinks	= true,		 
	urlcolor     = blue,	 
	linkcolor	 = purple, 	 
	citecolor    = violet    
}
\usepackage[capitalize]{cleveref}

\newcommand{\Regret}{\kR}

\newcommand{\EE}[1]{\bE\left[#1\right]}

\newcommand{\Prob}[1]{\bP\left\{#1\right\}}

\renewcommand{\phi}{\varphi}
\renewcommand{\epsilon}{\varepsilon}

\newcommand{\al}[1]{ \begin{align} #1  \end{align}}
\newcommand{\eq}[1]{ \begin{equation} #1  \end{equation}}
\newcommand{\als}[1]{ \begin{align*} #1  \end{align*}}
\newcommand{\eqs}[1]{ \begin{equation*} #1  \end{equation*}}

\newcommand{\el}{\end{flushleft}}
\newcommand{\bl}{\begin{flushleft}}

\newcommand{\ceil}[1]{ \left\lceil #1 \right\rceil }

\newcommand{\argmax}{\arg\!\max}

\newcommand{\bE}{\mathbb{E}}

\newcommand{\bP}{\mathbb{P}}

\newcommand{\kR}{\mathfrak{R}}

\theoremstyle{plain}

\newtheorem{lem}{Lemma}

\newtheorem{rem}{Remark}

\usepackage{dblfloatfix}

\begin{document}

\title{Exploiting Side Information for Improved Online Learning Algorithms in Wireless Networks}

\author{\IEEEauthorblockN{Manjesh K. Hanawal}
\IEEEauthorblockA{\textit{MLiONS Labs, IEOR} \\
\textit{IIT Bombay, Mumbai, India 400076}\\
mhanawal@iitb.ac.in}

\and
\IEEEauthorblockN{Sumit J. Darak}
\IEEEauthorblockA{\textit{Algorithms to Architectures Lab, ECE}\\
 \textit{IIIT Delhi, New Delhi, India 110020}\\
sumit@iiitd.ac.in}
}

\maketitle

\begin{abstract}

In wireless networks, the rate achieved depends on factors like level of interference, hardware impairments, and channel gain. Often, instantaneous values of some of these factors can be measured, and they provide useful information about the instantaneous rate achieved. For example, higher interference implies a lower rate. In this work, we treat any such measurable quality that has a non-zero correlation with the rate achieved as side-information and study how it can be exploited to quickly learn the channel that offers higher throughput (reward). When the mean value of the side-information is known, using control variate theory we develop algorithms that require fewer samples to learn the parameters and can improve the learning rate compared to cases where side-information is ignored. Specifically, we incorporate side-information in the classical Upper Confidence Bound (UCB) algorithm and quantify the gain achieved in the regret performance. We show that the gain is proportional to the amount of the correlation between the reward and associated side-information. We discuss in detail various side-information that can be exploited in cognitive radio and air-to-ground communication in $L-$band. We demonstrate that correlation between the reward and side-information is often strong in practice and exploiting it improves the throughput significantly. 

\end{abstract}

\section{Introduction}
\label{sec: Intro}

Rate achievable in a wireless network is a function of the signal to interference noise ratio given as $\frac{|h|^2P}{I_0+\eta}$, where
$P$ is the transmission power in watts, $h$ is the instantaneous channel gain, $I_0$ is the total interference power in watts, and $\eta$ is the total noise power in watts at the receiver. Interference, channel gain, and noise can be stochastic and their realization may not be known before transmission. However, in many wireless systems, it may be possible to measure instantaneous values of some of these quantities and can be used as side-information (SI). In this work, we exploit side-information to improve the performance of learning algorithms in wireless networking to identify best quality (rate/throughput/success) channels. 

Multi-armed Bandit (MAB) framework is widely used for learning in various wireless setting, like Millimeter waves communication \cite{JSAC2017_MillimeterWaves}, Cognitive radio networks \cite{JSTSP2011_CognitiveRadioSurvey_BeibeiRay}, and ad-hoc networks \cite{JNCA2014_AdHocNetworkSurvey_AlSultan}. In all these settings, often the goal is to learn the channels or beam directions that give the highest throughput. The learning algorithms in these settings use the rate achieved or transmission success at the receiver as feedback to identify the best channels. However, in these networks, often other SI, having a non-zero correlation with the received rate, could be exploited to improve the performance of MAB algorithms. 

Wireless networks protocols provide provision to measure inference and channel gain \cite{NRbulletsbook}[Chap. 3], \cite{5G_interference_measurement}. A transmitter can sense the channel to measure the level of interference before the information is transmitted in a slot. If measured interference at the transmitter in a slot is higher, possibly the receiver may also experience higher interference in that slot.  Of course, the receiver may experience inferences from other sources that the transmitter may not see, still, the rate at the receiver can be correlated with the interference level seen at the transmitter. Also, in the networks where uplink and downlink are used for data transmission using time division duplexing \cite{NRbulletsbook}[Chap. 4], the transmitter can measure the channel gain using pilots in the uplink transmissions. This channel gains will have a strong correlation with the rate received on the downlink and can be used as side-information.

In this work, we assume that the mean of the side-information is known, or can be estimated to good accuracy. This assumption may appear strong and overly restrictive, but it is possible to get this information in wireless networks; the mean of side-information could be available from the measurements during the network planning and deployment phase or could be obtained by offline measurements to good accuracy. When the mean values are known, the side-observations can be used to get a better estimate of the quality of channels that has smaller variance and sharper confidence bounds. Specifically, we combine the rate and 'centered' side-information samples with an 'optimal' weight and use the new samples to estimate the channel rates. The optimal weight depends on the correlation coefficient between the rate and the side-information and may not be known, but can be estimated.

We build on the Upper Confidence Bound (UCB) algorithm using the new estimators that exploit the side-information. The samples used for the new estimator are correlated and this brings out significant challenges in the analysis of the new UCB algorithm as standard concentration bounds cannot be used. For the special case where reward and side-information are jointly Gaussian, we provide performance guarantees for the new UCB algorithm leveraging the results from control variate theory. The new algorithm is shown to offer better performance than vanilla UCB that does not exploit side-information and the magnitude of improvement depends on the correlation between the reward and side-information for each channel. For the general distributions, we use one of the popular re-sampling techniques named Splitting to leverage the side-information and improve the performance of the UCB algorithm. Our contributions can be summarized as follows:

\begin{itemize}
\item We develop a new framework in \cref{sec: Setting} which exploits the side-information in wireless networks to improve the performance of learning algorithms. 
\item For the case where the reward and side-information are jointly Gaussian, we develop an algorithm named UCB with Side-Information (UCBwSI) in\cref{sec: Algorithm}. We show that its regret is better by a factor at least $(1-\rho^2)$ in \cref{thm:regretBound}, where $\rho=\min_{i}\rho_i$ is the minimum correlation coefficient between rate and side-information across all the arms.
\item For general distributions, we give an algorithm named UBC with Side-Information using Splitting (UCBwSI-Split) in \cref{sec: Gen_Distri}. This algorithm is based on the popular re-sampling method named Splitting.
\item We take Cognitive Radio Networks and $L$-band Digital Aeronautical Communication System (LDACS) as case studies in \cref{sec: CognitiveRadio} and \cref{sec: LDACS} to demonstrate various side-information available in wireless networks. We simulate these networks over a wide range of realistic parameters and validate the performance improvement achieved by the new algorithms.
\end{itemize}

\subsection{Related Work}
Multi-armed Bandit (MAB) framework have been used extensively to develop learning algorithms in various wireless setups, like cognitive radio networks \cite{JSAC11_DistributedLearning_Anadakumar, TIT14_DecentralizedLearning_KalthilNayyarJain, TCNS2018_DecentralizedLearning_KalthilNayyarJain, Infocom2019_DistributedLearning_TibrewalPatchalaHanawal}, Energy harvesting networks \cite{MCL2021_MABEnergyHavesting,JSAC2015_MABEnergyHarvesting,TIT2018_OnlineLearningEnegyHarvesting}, Millimeter Wave communications \cite{INFOCOM2020_MAMBA,INFOCOM2020_OnlineBayesianLearningmmWave}, and 5G cellular networks \cite{VTC2018_MAB5G,TCOM2021_MABNOMA,SSPW2021_MABNOMA}. We focus on the stochastic setting where classical bandit algorithms like UCB1 \cite{ML02_auer2002finite}, KL-UCB \cite{COLT11_garivier2011kl}, and Thompson sampling \cite{AISTATS2013_RegretBoundsForThomsonSampling_AgarwalGoyal} are applied to networks by suitably mapping actions (channels, modulations schemes, power level) to arms and quantity of interests (success rate, throughput) to rewards to obtain learning algorithms. As our focus is on better adapting the MAB setting to networks, in the following we discuss the works that incorporate properties of the wireless networks or any specific network information to improve learning performance. For details on various MAB setup in wireless networks we refer to \cite{Book_OnlineLearningNetworks, WirelessCommunications2020_MAB_Survey,WirelessCommunication2016_MAB5G}. 

\noindent
In wireless networks,  if the transmission is successful at a rate, transmission at a lower rate is more likely to be successful, and if the transmission fails at a rate, use of any higher rate is more likely to fail. This gives rise to {\it unimodal} property and is exploited in MAB algorithms for rate sampling in Wi-Fi \cite{INFOCOM2014_OptimalRateSampling_Combes, TMC2019_OptimalRateSampling_Combes}. Unimodal property or correlation structure also arises in beam alignment in millimeter-wave communication \cite{INFOCOM2018_MABmmWave_HashemiSbharwalShroff,TWC2019_FastmmWave}: if matching between a pair of transmitter and receiver beams with a certain misalignment is successful, any pair with lower misalignment is more likely to be successful, and if a pair has poor matching, then all the pairs with larger misalignment are more likely to be poorer. By exploiting the unimodality property, sub-optimal options can be quickly eliminated resulting in faster MAB algorithms.  

\noindent
Dummy probe packets can be used in networks to get additional information about the success probability of a transmission rate. Such probes result in throughput loss but provide additional information about throughputs from different rates. \cite{MACS2018_MABSideObservations_YunProutiere} demonstrate that exploitation of such {\it additional information} better rate selection in 802.11-like wireless systems. When the interference graph of a network is known, the authors in \cite{ITW2013_MABSpectrumBandit_LeLargeProtiere} exploit the graph structure to develop efficient MAB algorithms to maximize the network throughput.
A transmitter can have contextual information like received signal strength, the cell area, user location, average number of devices in the channel. 
Authors in \cite{NetworkMeets_ContextualMAB,ICOIN2021_ContextualBanditIndustry4.0} exploit such contextual information to improve link selection in cellular networks.

Our work differs from the previous as we do not require any structure in the reward or seek additional information. We utilize readily available side-information to improve the performance of MAB algorithms. The side-information in our work can also be used as contextual information, but we do not base channel selection on this as it may not be available for all the channels before they are selected. 

We use side-information to obtain unbiased point-estimates that have smaller variance and confidence intervals using control variate theory  \cite{JORS85_james1985variance, OR90_nelson1990control}. Control variate method is popular various reduction technique used extensively in Monte-Carlo simulations \cite{MS82_lavenberg1981perspective,OR82_lavenberg1982statistical,EJOR89_nelson1989batch}. The new estimators are used in UCB algorithm to get improved performance. To the best of our knowledge, leveraging control variate theory to improve performance online algorithms for wireless networks is new.

\section{Problem Setup}
\label{sec: Setting}

\begin{figure}[!t]
\centering
\vspace{-0.2cm}
    \includegraphics[width=0.9\linewidth]{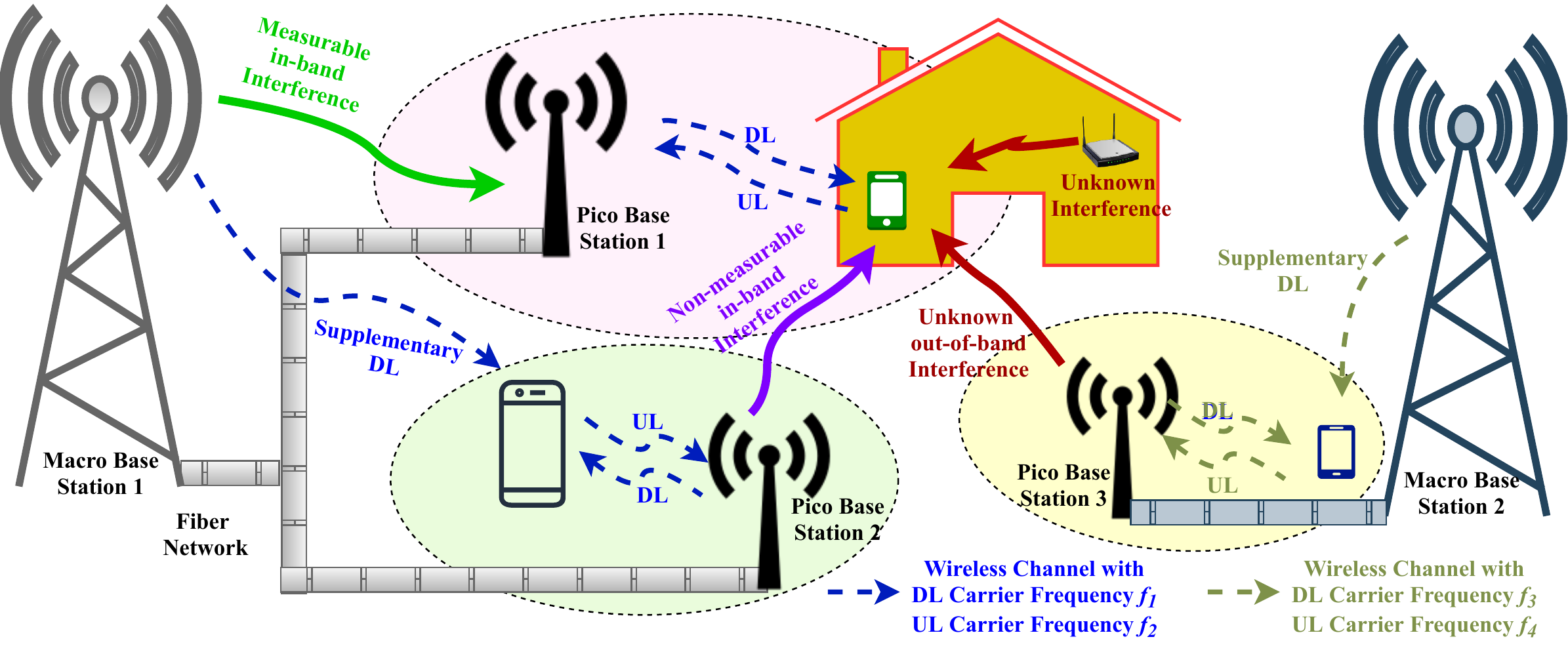}
    \vspace{-0.2cm}
    \caption{An heterogeneous network consisting of macro and picocells. UEs see interference from other base stations and networks. Some of these could be measurable and can be used side-information}
    \label{fig:Setup}
    \vspace{-2mm}
\end{figure}

Consider a heterogeneous network consisting of macrocells and picocells as shown in \cref{fig:Setup}. A user equipment (UE) in a picocell communicates with its picocell base station through uplink and downlink channels and with a macro base station through a supplementary downlink channel. A UE gets interference (both in-band and out-of-band) from neighboring pico base stations, macro base stations, and other sources like wi-fi networks. Consider pico base station $1$.  It can measure interference level from the macro base station $1$ (through fiber back-haul or sensing), but not the other inference experienced at the UE served by it. Then, the base station $1$ can use the measured interference as side-information and exploit it to select channel for UE. Other network setups like Cognitive Radio and LDACS are discussed in detail in Sections \ref{sec: CognitiveRadio} and \ref{sec: LDACS}. In the following, we abstract the network settings and set it up as a learning problem where the the aim of the learner is to identify channel offering highest throughput\footnote{If the learns has to learn the best Modulation and Coding Scheme (MCS), the same setup can be used with the set of MCS as actions.}.

Let the set of channels be denoted by $[K] \doteq \{1, 2, \ldots, K\}$. The rate achieved on a channel $i\in [K]$ depends on various stochastic quantities denoted as $\{W_i, W_{-i}\}$ where $W_i$ denote the quantity that can be observed by the learner and $W_{-i}$ denote the set of all other that are not observed. We refer to $W_i$ as side-information for channel $i$. The rate achieved or throughput on channel $i$ is denoted as $X_i$ and is an unknown function of $\{W_i, W_{-i}\}$, denoted as $X_i=f_i(W_i, W_{-i})$. For example, consider that transmitter can observe the interference on the channel $i$, and other factors like channel gains ($h_i$), receiver noise $(\eta_i)$, interference at the receiver $(I_i)$ are not observed.  Then $W_i$ is the interference at the transmitter and other quantities constitute $W_{-i}=(h_i,\eta_i,I_i)$ and the rate on channel $i$ as a function of SINR can be given by
\begin{equation}
\label{equ: Rate1}
 X_i=f_i(W_i,W_{-i})=f\left( 
\frac{|h_i|^2P}{|h_i|^2W_i+I_i+\eta_i}\right),   
\end{equation}
where $P$ is the power transmitted.
If channel gain is observed at the transmitter, but total interference at the receiver ($I_i$) and a receiver noise $(\eta_i)$ are not observed, then $W_{-i}=(I_i, \eta_i)$ and rate can be expressed as 
\begin{equation}
\label{equ: Rate2}
X_i=f_i(W_i, W_{-i})=f\left(
\frac{W_i P}{I_i+\eta_i}\right).
\end{equation}
In the following, we assume the availability of one side-information for each arm\footnote{When multiple side-information are available for the arms, we can extend the analysis straightforwardly. See \cref{rem: Extensions} later}.

In the following we refer to the channels as arms and rate received/throughput as reward. In round $t$, let $X_{t,i}$ and $W_{t,i}$ denote reward and side-information from arm $i$. The sequence $(X_{t,i}, W_{t,i})_{t\geq 1}$ is assumed to be independently and identically distributed (i.i.d.) across the time for each $i\in [K]$.  For the $i$-th channel, we denote the mean and variance of reward as $\mu_i=\mathbb{E}[X_{t,i}]$ and $\sigma_i^2=\mbox{var}(X_{t,i})$. The mean value of side-information is denoted as $\omega_i=\mathbb{E}[W_{t,i}]$ and the correlation coefficient between reward and side-information pair $(X_{t,i}, W_{t,i})$ is denoted as $\rho_i$. We assume that learner knows only $\{\omega_i\}_{i\in [K]}$ and none of other quantities.

In our setup, the learner observes the reward and side-information from the selected arm in each round. The goal is to learn the arm with highest mean, i.e., $i^*=\arg \max_{i \in [K]} \mu_i$ quickly. The learner plays the arms sequentially and in each round, the learner has to decide which arm to play based on the past observations. We aim to learn policies that accumulate reward as close as possible to the optimal policy which is to play arm $i^*$ in each around.  We measure the performance of any policy by comparing its cumulative reward with that of the optimal policy.  Specifically, the performance of a policy  that selects arm $I_t$ in round $t$ is  measured in terms of expected regret defined as follows:
\begin{equation}
	\label{eqn:regret}
	\Regret_T = T\mu_{i^\star} - \EE{\sum_{t=1}^{T} X_{t,I_t}}.
\end{equation}
A good policy should have sub-linear expected regret, i.e., ${\Regret_T}/T \rightarrow 0$ as $T \rightarrow \infty$. Note that the above regret definition is the same as that without the side-information. Our goal is to achieve smaller regret by exploiting the side-information compared to the case when no such information is available. To this effect, we use side-information to derive sharper mean estimators so that the learner can achieve a exploration-exploitation trade-off and start playing the optimal arm more frequently earlier.  

\section{Algorithm for Gaussian Distributions}
\label{sec: Algorithm}

In this section, we focus on the case where reward and side-information are jointly  Gaussian for all arms.  The case of arbitrary distribution is considered in the next section.  We first discuss how to use side-information to construct a better point-estimator with a smaller variance. We then develop a UCB based algorithm using the new estimator.

For all $i\in [K]$, let $\hat{\mu}_{s,i}$ and $\hat{\omega}_{s,i}$ denote the sample mean estimator of reward and side-information, i.e.,
\[\hat{\mu}_{s,i} = \frac{1}{s} \sum_{r=1}^s X_{r,i}  \quad \mbox{and} \quad \hat{\omega}_{s,i} = \frac{1}{s} \sum_{r=1}^s W_{r,i}. \]
We generate a new sample mean estimator for a reward using samples that are obtained by linearly combining reward and side-information samples. This linear combination is  motivated by the linear control variate technique. A brief discussion on the control variate theory is given in Appendix A. Let $(X_{t,i},W_{t,i})$ be the reward and side-information sample pair from arm $i$ in round $t$. A new sample is obtained as follows:
\eq{
    \label{equ:observedSample}
    \bar{X}_{t,i} = X_{t,i} + \beta^*_i(\omega_i - W_{t,i}),
}
where  $\beta_i^*$ is a constant. Let  $\hat\mu_{s,i}^c$ denote the point-estimator using $s$ new samples given as 
\eq{
	\label{equ:estAvgReward}
	\hat\mu_{s,i}^c = \frac{\sum_{r=1}^s \bar{X}_{r,i}}{s}=\hat\mu_{s,i} + \beta^*_{i}(\omega_i - \hat\omega_{s,i}).
}
Notice that 	$\hat\mu_{s,i}^c$ is an unbiased estimator. Further, when  $\beta_i^*$ is set appropriately, the variance of  $\hat\mu_{s,i}^c$ is smaller than that of $\hat\mu_{s,i}$ as given by the following lemma. 

\begin{lem}\cite{MS82_lavenberg1981perspective}
\label{lem: OptimalBeta}
Let  $Cov(X_{i},W_{i})$ denotes the covariance between reward and side-information pair $(X_{i},W_{i})$ and $\text{Var}(W_{i})$ denotes the variance of  side-information from the $i$-th arm. Set $\beta_i^*=\text{Cov}(X_{i},W_{i})/\text{Var}(W_{i})$, Then, 
\[\text{Var}(\hat\mu_{s,i}^c)=(1-\rho_i^2)\text{Var}(\hat{\mu}_{s,i})=(1-\rho_i^2)\sigma_i^2/s.\]
\end{lem}
\noindent
To realize the improvement in the variance of  $\hat\mu_{s,i}^c $,  we need to know the covariance between reward and side-information and variance of the side-information both of which may not be known a priori. But they can be estimated and we  replace  $\beta_i^*$ by its estimate given as follows
\eq{
	\label{equ:estBeta}
	\hat\beta^*_{s,i}= \frac{\sum_{r=1}^{s} (X_{r,i} - \hat\mu_{s,i})(W_{r,i} - \omega_i)}{\sum_{r=1}^{s}(W_{r,i}-\omega_i)^2}.
}
We added subscript $s$ to the estimate of $\beta^*_i$ to make dependency on the number of samples explicit.
Replacing $\beta^*_i$ by its estimate in  (\ref{equ:estAvgReward}),  the new point-estimate for mean reward of arm $i$ is given as
\eq{
	\label{equ:NewestAvgReward}
	\hat\mu_{s,i}^c = \hat\mu_{s,i} + \hat{\beta}^*_{s,i}(\omega_i - \hat\omega_{s,i}).
}
The new estimator $\hat\mu_{s,i}^c$ is a  mean of non i.i.d. samples as  $\hat{\beta}^*_{s,i}$ depends on the past $s$ samples and we cannot apply \cref{lem: OptimalBeta} to estimate variance of $\hat\mu_{s,i}^c$. It is in general hard to calculate it for an arbitrary distribution due to the dependency introduced by $\hat{\beta}^*_{s,i}$ . However, for the special case where reward and side-information are jointly Gaussian, one can estimate the variance of $\hat\mu_{s,i}^c$ as shown in the following Proposition
.
\begin{restatable}{prop}{varEst}
	\label{prop:varEst}
	Let the reward and side-information pair for each arm be jointly Gaussian. For $s$ reward and side-information sample pairs of arm $i$, define
$$S^2=\frac{\sum_{r=1}^{s} \left(\bar{X}_{r,i}-\hat{\mu}_{s,i}^c\right)^2}{s-2}. \quad \mbox{Then}$$
$$\hat{\nu}_{s,i}=S^2 \left(\frac{1}{s}-\frac{(\sum_{r=1}^s(W_{r,i}-\omega_{i}))^2}{s^2\sum_{r=1}^{s}(W_{r,i}-\hat{\omega}_{s,i})^2}\right)^{-1}$$
 is an unbiased variance estimator of $\hat{\mu}^c_{s,i}$, i.e., $\mathbb{E}[\hat{\nu}_{s,i}]=\text{Var}(\hat{\mu}^c_{s,i})$.
\end{restatable}
\noindent
For Gaussian distributed reward with variance $\sigma_i^2$, Eq. (\ref{equ:observedSample}) can be rewritten as  $ X_{t,i} = \mu_i+ \beta^*_i(\omega_i - W_{t,i})+ \epsilon_{t,i}$ where $\epsilon_{t,i}$ is a zero Gaussian random variable with variance $(1-\rho_i^2)\sigma_i$. The proof exploits the fundamental properties of the linear regression. Detailed proof of the Proposition is given in  \cite{OR90_nelson1990control}[Thm 1]. With an unbiased variance estimator for $\text{Var}(\hat{\mu}^c_{s,i})$, we can construct the tight confidence intervals on the reward mean estimator as given by our next result.
\begin{restatable}{prop}{confProb}
    \label{prop:confProb}
   For each arm, let reward and side-information be jointly Gaussian. Then for all $i\in [K]$ we have
    \eqs{
        \Prob{|\hat\mu_{s,i}^c - \mu_i| \ge V_{t,s}^{(\alpha)}\sqrt{{\hat\nu_{s,i}}}} \le 2/t^\alpha,
    }
where $V_{t,s}^{(\alpha)}$ denote $100(1-1/t^\alpha)^{\text{th}}$ percentile value of the $t-$distribution with $s-2$ degrees of freedom and   $\hat\nu_{s,i}$ is an unbiased estimator for  variance  of $\hat{\mu}^c_{s,i}$.
\end{restatable}
The proof follows along similar lines in \cite{OR90_nelson1990control}[Thm. 1] after setting appropriate percentile values for the $t$-distribution.  The above concentration bound is analogous to Hoeffding inequality, which shows that the deviation of the estimate of mean reward obtained by samples $\{\bar{X_{t, i}}\}$ around the true mean decays very fast. Note that the  deviation factor $V_{t,s}^{(\alpha)}\sqrt{{\hat\nu_{s,i}}}$ depends on the variance of the estimator $(\hat{\mu}^c_{s,i})$ which guarantees sharper confidence intervals. As we will see, these confidence terms are smaller by a factor $(1-\rho_i^2)$ compared to the case of no side-information use. We note that this improvement in confidence bound is achieved without requiring any additional samples. Hence algorithms using the new confidence bounds are expected to learn the optimal arm faster. 

\subsection{Algorithm: \ref{alg:UCBwSI}}
Let $N_i(t)$ denote the number of times $i$-th arm  is selected till round $t$ and $\hat\nu_{N_i(t),i}$ be the variance of mean reward estimator $(\hat\mu_{N_i(t),i}^c)$ for arm $i$. Following Proposition \ref{prop:confProb}, we define optimistic upper confidence bound for estimate of mean reward of arm $i$ as follows: 
\begin{equation}
	\label{equ:UCB}
	\text{UCB}_{t,i} = \hat\mu_{N_i(t),i}^c + V_{t,N_i(t),1}^{(\alpha)}\sqrt{{\hat\nu_{N_i(t),i}}}.
\end{equation}
Treating the above value as indices of arms, we develop an algorithm named UCB with Side-Information (\ref{alg:UCBwSI}). Its pseudo-code is given below.
\begin{algorithm}[!ht] 
	\renewcommand{\thealgorithm}{UCBwSI}
	\floatname{algorithm}{}
	\caption{UCB with Side Information Algorithm}
	\label{alg:UCBwSI}
	\begin{algorithmic}[1]
		\STATE \textbf{Input:} $K,  ~\alpha$
		\STATE Play each arm $i \in [K]$ $4$ times
		\FOR{$t=4K+1, 4K + 2, \ldots, $}
			\STATE $\forall i \in [K]:$ compute UCB$_{t-1,i}$ as given in \cref{equ:UCB}
			\STATE Play $I_t = \argmax\limits_{i \in [K]}$ UCB$_{t-1,i}$
			\STATE Observe $X_{t,I_t}$ and $W_{t,I_t}$. Increment $N_{I_t}(t)$ by one and re-estimate $\hat{\beta}^*_{N_{I_t}(t),I_t}$, $\hat\mu_{N_{I_t}(t),{I_t}}^c$ and $\hat\nu_{t,N_{I_t}(t)}$
		\ENDFOR
	\end{algorithmic}
\end{algorithm}
The algorithm works as follows: It takes $K$ and $\alpha>1$ as inputs. $\alpha$ determines the trades-off between exploration and exploitation. At the start, each arm is played $4$ times to ensure that the sample variance $\hat\nu_{s,i}$ is well defined (see Prop. \ref{prop:varEst}). In round $t>4K$, \ref{alg:UCBwSI} computes UCB index of each arm as in \cref{equ:UCB} and selects the arm with the highest UCB index. The arm selected is denoted as $I_t$. After playing arm $I_t$, the reward $X_{t, I_t}$ and side-information $W_{t, I_t}$ are observed. Then, the value of $N_{I_t}(t)$ is updated and $\hat{\beta}^*_{N_{I_t}(t),I_t}$, $\hat\mu_{N_{I_t}(t),{I_t}}^c$ and $\hat\nu_{t,N_{I_t}(t)}$ are re-estimated. The same process is repeated till the end. We note that \cref{alg:UCBwSI} uses variance of the estimator to define the UCB index whereas other variance estimation based algorithm like UCBV \cite{TCS09_audibert2009exploration} user variance of the reward.

\subsection{Analysis of \ref{alg:UCBwSI}}
Our regret bound is based on the following classical result from control variate theory.
\begin{restatable}{prop}{estProp}\cite{OR90_nelson1990control}[Thm. 1]
    \label{prop:estProp}
    Let reward and side-information are jointly Gaussian for all the arms. The estimate in (\ref{equ:NewestAvgReward}) obtained with $s>3$ reward and side-observations samples satisfies the following properties:
    \als
    {
     \EE{\hat\mu_{s,i}^c} = \mu_i \text{ and }
         \text{Var}(\hat\mu_{s,i}^c) = \frac{s-2}{s-3}(1-{\rho^2_i})\text{Var}(\hat\mu_{s,i}). 
    }
\end{restatable}
\noindent
Following is our main result which upper bounds regret of \ref{alg:UCBwSI}. Its proof is given in Appendix C.
\begin{restatable}{thm}{regretBound}
	\label{thm:regretBound}	
	Let the \ref{alg:UCBwSI} is run with  $\alpha=2$.  Then the  regret of \ref{alg:UCBwSI} in $T$ rounds is upper bounded by
	\eqs{
		\Regret_T \le 8\sum_{i \ne i^\star} \hspace{-0.5mm} \left( \frac{(V_{T,T}^{(2)})^2C_{T,i} (1-\rho_i^2)\sigma_i^2}{\Delta_i} + \frac{\Delta_i\pi^2}{3} + \Delta_i\right),
	}
where $\Delta_i = \mu_{i^\star} - \mu_i$ is the sub-optimality gap for arm $i \in [K]$ and 
$C_{T,i}=\mathbb{E}\left [{\left(V_{T,N_i(T)}^{(2)}/V_{T,T}^{(2)}\right)^2}\right]$.
\end{restatable}
\noindent
Unfortunately, we cannot directly compare our bound with that of UCB1, UCB1-NORMAL \cite{ML02_auer2002finite}, and UCBV \cite{TCS09_UCBV_Audibert} as $V_{T, T}^{(2)}$ do not have closed-form expression. $V_{T, T}^{(2)}$ denotes $100(1-1/T^2)^\text{th}$ percentile of student's $t$-distribution with $T-2$ degrees of freedom. The value of $V_{T,T}^2$ can be numerically shown to be upper bounded by $3.726\log(T)$. $C_{T,i}$ also do not have close form expression, but $C_{T,i}\rightarrow 1$ as $T\rightarrow \infty$. When each arm is explored sufficiently ($N_i(T)\sim 40$), one can numerically verify that $C_{T,i}\leq 1.5$.

\begin{rem}
	\label{rem: Extensions}
The above results can be slightly generalized to the case where rewards are Gaussian but need not by jointly Gaussian with side-information as shown in \cite{OR90_nelson1990control}[Thm. 2]. Also,  the analysis can be extended when the arms have more than one side-information. More details are in Appendix B.
\end{rem}
\begin{rem}
\label{rem: Extensions2}
If the mean side-information is unknown, the algorithm can use an approximate mean or estimating the mean from samples as explained in  	\cite{IIE2001_BiasedControlVariate_Bruce} and   \cite{IIE2012_ControlVariate_Raghu}.
\end{rem}

\section{General Distribution}
\label{sec: Gen_Distri}

When reward and side-information have arbitrary distribution, 
$\hat\mu_{s,i,q}^c$ is no  more guaranteed to be Gaussian and unbiased estimator. Therefore, we cannot obtain confidence intervals using properties of $t$-distributions as done in \cref{prop:confProb}. However, we can use re-sampling methods such as Jackknifing, Splitting to reduce  bias of the estimators and develop confidence intervals that hold approximately or asymptotically.  Below we discuss an algorithms based on the  Splitting method.

\subsection{Algorithm with Splitting aAproach}
\label{ssec:splitting}
Splitting is a re-sampling technique that splits the correlated observations into two or more groups, compute an estimate of $\boldsymbol{\beta}_i^*$ from each group, then exchange the estimates among the groups. We consider the  form of splitting discussed in \cite{OR90_nelson1990control} [Sec. 6] where $s$ groups are formed. The $j$the group, $j\in [s]$, is obtained by dropping the $j$-th reward and side-observation pair. From each group $j\in [s]$, $\beta^*_i$ is estimated and denoted $\hat{\beta}_{(s-1),i}^{*-j}$. The new samples are then calculated as 
\eqs{
    \bar{X}_{j,i}^{\text{S}} = X_{j,i} + \hat{\beta}_{(s-1),i}^{*-j}(\omega - W_{j,i}), ~~\forall~ j \in [s],
}
The point estimator for splitting method  is $\hat\mu_{s,i}^{c, \text{S}} = \frac{1}{s} \sum_{j=1}^{s}  \bar{X}_{j,i}^{\text{S}}$ and its sample variance is $\hat\nu_{s,i}^{\text{S}} = (s(s-1))^{-1}$ $\sum_{j=1}^s ( \bar{X}_{j,i}^{\text{S}} - \hat\mu_{s,i}^{c, \text{S}})^2$. Then $\hat\mu_{s,i}^{c, \text{S}} \pm t_{\alpha/2}(n-1)\hat\nu_{s,i}^{\text{S}}$ gives an approximate confidence interval \cite{OR90_nelson1990control}. Using this we define the optimistic upper confidence bound for mean reward as
\eq{
    \label{equ:genDistUCB}
	\text{UCB}_{t,i}^{\text{S}} = \hat\mu_{N_i(t),i}^{c, \text{S}} + V_{t,N_i(t)}^{\text{S},\alpha}\sqrt{\hat\nu_{N_i(t),i}^{\text{S}}}.
}
where $V_{t, s, q}^{\text{G}, \alpha}$ is the $100(1-1/t^\alpha)^{\text{th}}$ percentile value of the $t-$distribution with $s-1$ degrees of freedom. We can use the above UCB indices to get an algorithm for the general distribution case. We refer to the algorithm as UCB with Side-information and Splitting (UCBwSI-Split). Its pseudo code is given in  \cref{alg:UCBwSI-Split}.
 
\begin{algorithm}[!h] 
	\renewcommand{\thealgorithm}{UCBwSI-Split}
	\floatname{algorithm}{}
	\caption{UCB with Side Information Using Splitting}
	\label{alg:UCBwSI-Split}
	\begin{algorithmic}[1]
		\STATE \textbf{Input:} $K,  ~\alpha>1$
		\STATE Play each arm $i \in [K]$ $4$ times
		\FOR{$t=3K+1, 3K + 2, \ldots, $}
		\STATE $\forall i \in [K]:$ compute $\text{UCB}^S_{t-1,i}$ as given in \cref{equ:genDistUCB}
		\STATE Play $I_t = \argmax\limits_{i \in [K]}$ $\text{UCB}^S_{t-1,i}$
		\STATE Observe $X_{t,I_t}$ and $W_{t,I_t}$. Increment the value of $N_{I_t}(t)$
		\STATE Compute $\hat\beta^{*-j}_{(N_{I_t}(t)-1),{I_t}}$ and  $\bar{X}_{j,I_t}^S$ for all $j \in [N_{I_t}]$
		\STATE Compute $\hat{\mu}^{c,S}_{N_{I_t}(t),I_t}$ and $\hat\nu_{t,N_{I_t}(t)}$
		\ENDFOR
	\end{algorithmic}
\end{algorithm}

Since the optimistic upper bound defined in \cref{equ:genDistUCB} holds only asymptotically \cite{OR90_nelson1990control, NRL91_avramidis1991simulation}, we cannot use them to provide any finite time guarantees as done in Prop. \ref{prop:confProb} and Thm. \ref{thm:regretBound} for the Gaussian case.
the general case. We experimentally validate its performance in the next section.

\section{Application to Cognitive Radio Networks}
\label{sec: CognitiveRadio}
\begin{figure*}[!t]
\centering
    \includegraphics[scale=.5]{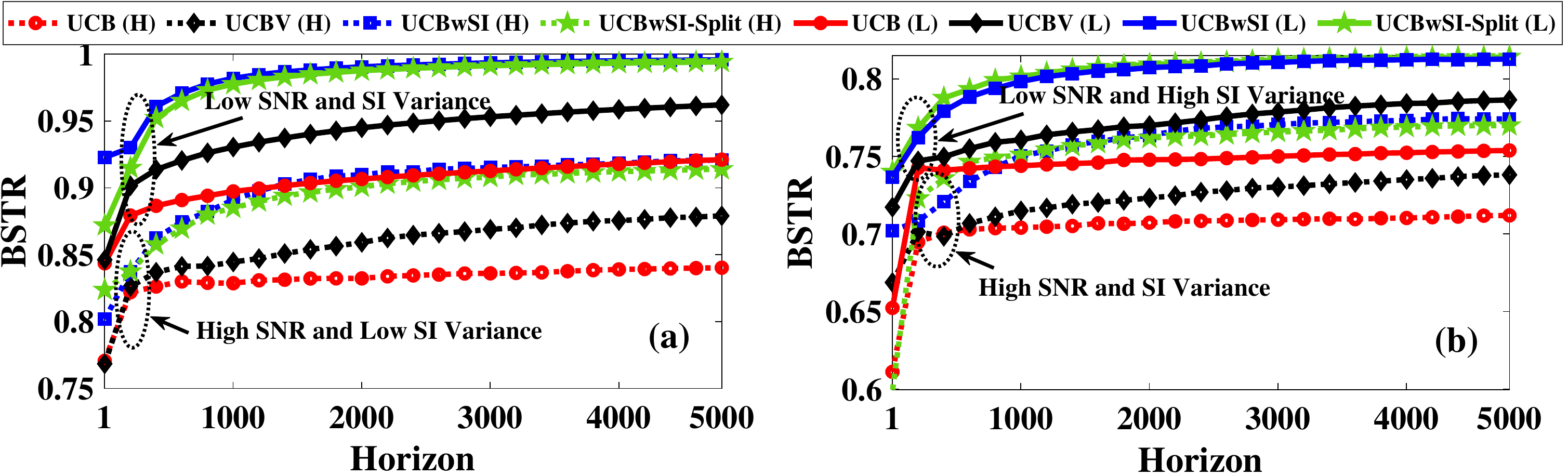}
    \caption{Effect of (a) Channel SNR variance, and (b) Interference power level variance on the achieved BSTR. Higher BSTR is preferred.}
    \label{fig:OFDM_var_comp}
\end{figure*}

We consider the cognitive radio (CR) networks in the unlicensed spectrum. A CR user (learner) aims to identify the wireless channel which offers a lower bit-error rate or higher Bit Successful Transmission Rate (BSTR) for given physical layer parameters such as coding rate, modulation scheme, and bandwidth. The power level of measurable interference at the transmitter is used as Side-Information (SI) on each channel. The performance of UCBwSI and UCBwSI-Split are compared against UCB1-Normal \cite{ML02_auer2002finite} (referred simply as UCB in the following) and UCBV \cite{TCS09_audibert2009exploration} both of which use variance estimate \footnote{We did not compare UCBwSI and UCBwSI-Split against KL-UCB and Thompson sampling as KL-UCB does not incorporate variance estimation and it is unclear how to incorporate side-information in Thompson sampling.}. For all the algorithms we compare measured BSTR performance instead of regret as expressions for computation of optimal BSTR (as in Eqs. \ref{equ: Rate1} \& \ref{equ: Rate2}) are not available for our realistic setup. All the experiments in this section assume $K=8$ wireless channels and orthogonal frequency division multiplexing (OFDM) based physical layer for CR users. Various parameters of the physical layers are summarized in Table~\ref{tab:CR_PHY}. Parameters such as the number of subcarriers (NSC) of 4096, subcarrier spacing (SCS) of \{15 KHz, 30 KHz\} and 64-QAM modulation scheme are similar to the 5G cellular physical layer.

\begin{table}[!h]
\centering
\caption{CR Physical Layer Parameters}
\vspace{-0.2cm}
\label{tab:CR_PHY}
\resizebox{0.47\textwidth}{!}{%
\begin{tabular}{|l|l|}
\hline
\textbf{Parameters} & \textbf{Value} \\ \hline
\textbf{No. of Subcarriers (NSC)} & \{128, 1024, 4096\} \\ \hline
\textbf{No. of Data Subcarriers} & \{116, 1012, 4088\} \\ \hline
\textbf{Sub-carrier Spacing (SCS)} & \{15 KHz, 30 KHz\} \\ \hline
\textbf{OFDM Symbols/Time Slot} & 2\\ \hline
\textbf{Time Slot Duration ($\mu$s)} & \{71.35, 35.68\} \\ \hline
\textbf{Modulation Scheme} & \{QPSK, 64-QAM\} \\ \hline
\textbf{(Bipolar) ADC Bits} & \{6, 16\} \\ \hline
\textbf{Wireless Channel} & AWGN with 5 KHz offset \\ \hline
\textbf{Wireless Channel Gain} & As per the distribution \\ \hline
\textbf{In-band Interference} & \begin{tabular}[c]{@{}l@{}}Narrowband spread over\\ few unknown subcarriers\end{tabular} \\ \hline
\textbf{Out-of-band Interference} & \begin{tabular}[c]{@{}l@{}}Wideband with non/partially \\  overlapping  freq. range\end{tabular} \\ \hline
\textbf{Interference Power Level} & As per the distribution \\ \hline
\end{tabular}%
}
\vspace{-0.2cm}
\end{table}

Two types of interference are considered: 1)~In-band Interference: The transmission bandwidth of the interferer overlaps with that of CR users. We consider high-power narrowband interference which affects only a few subcarriers of the CR users, 2)~Out-of-band Interference: Interferers and CR users transmit over non-overlapping but adjacent frequency bands. In both cases, the transmit power of the interferers can have a significant impact on the BSTR of the CR users. The wireless channel is modeled as a basic additive white Gaussian noise (AWGN) channel. This is done to exclude the performance degradation due to multi-path and doppler effects and non-ideal channel estimation methods. This means the achieved BSTR will mainly depend on the channel SNR, interference, and impairments in analog-front-end (AFE). Realistic wireless channel effects are considered in Section~\ref{sec: LDACS}.

We model frequency division duplexing (FDD) system based CR communication in which downlink channel SNRs are unknown to the transmitter but modelled as stochastic parameter. The mean of the measurable interference is known and its instantaneous value can be measured. Impairments due to AFE are fixed but unknown. The mean SNR distribution of wireless channels is set as $\mu^1~=~\{ 5,-1,3,-9,7,-2,18,-7\}$ with two types of variances: 1) Low variance, $\sigma^{1}_{L}~=~\{0.5,0.5,1,0.8,0.1,0.3,0.2,0.4\}$, and 2) High variance, $\sigma^{2}_{H}~=~\{2,2,2,2,2,2,2,2\}$. Among various measurable, non-measurable and unknown interference, we consider the measurable interference as the SI with the mean received power distribution at the CR transmitter as $w_1=\{1.7,0.2,-3,-0.9,-0.4,1,-0.6,1\}$ with two types of variances: 1) Low variance, $\sigma^{w1}_{L} = \{0.2,0.4,0.3,0.2,0.3,0.1,0.4,0.7\}$, and 2) High variance, $\sigma^{w1}_{H} = \{2,2,2,2,2,2,2,2\}$. Each experiment runs for $5000$ time slots and results are averaged over $5$ iterations.

\noindent
\textbf{Effect of variance on BSTR:} In  Fig.~\ref{fig:OFDM_var_comp}, we compare the performance of the four candidate algorithms in terms of BSTR at different time instant of the horizon. We consider four combinations: 1) Low SNR and low SI variance (LL): $\{\mu^1,\sigma^1_{L},w_1,\sigma^{w1}_L\}$, 2) High SNR and low SI variance (HL): $\{\mu^1,\sigma^1_{H},w_1,\sigma^{w1}_L\}$, 3) Low SNR and high SI variance (LH): $\{\mu^1,\sigma^1_{L},w_1,\sigma^{w1}_H\}$, and 4) High SNR and high SI variance (HH): $\{\mu^1,\sigma^1_{H},w_1,\sigma^{w1}_H\}$. In all cases, UCBwSI and UCBwSI-Split outperform UCB and UCBV.  Furthermore, Fig.~\ref{fig:OFDM_var_comp} (a) and (b) shows that the low SNR variance leads to improved BSTR for low and high SI variance, respectively, due to faster learning of the channel statistics. Furthermore, for a given SNR variance, low SI variance offers higher BSTR. 
On an average, the BSTR of UCBwSI is higher by 0.8 and 0.5 than UCB and UCBV, respectively. This corresponds to 0.5 and 0.3 mega bits per seconds (MBPS) improvement in throughput in each time slot, over UCB and UCBV, respectively.

\begin{figure}[!b]
    \includegraphics[scale=.32]{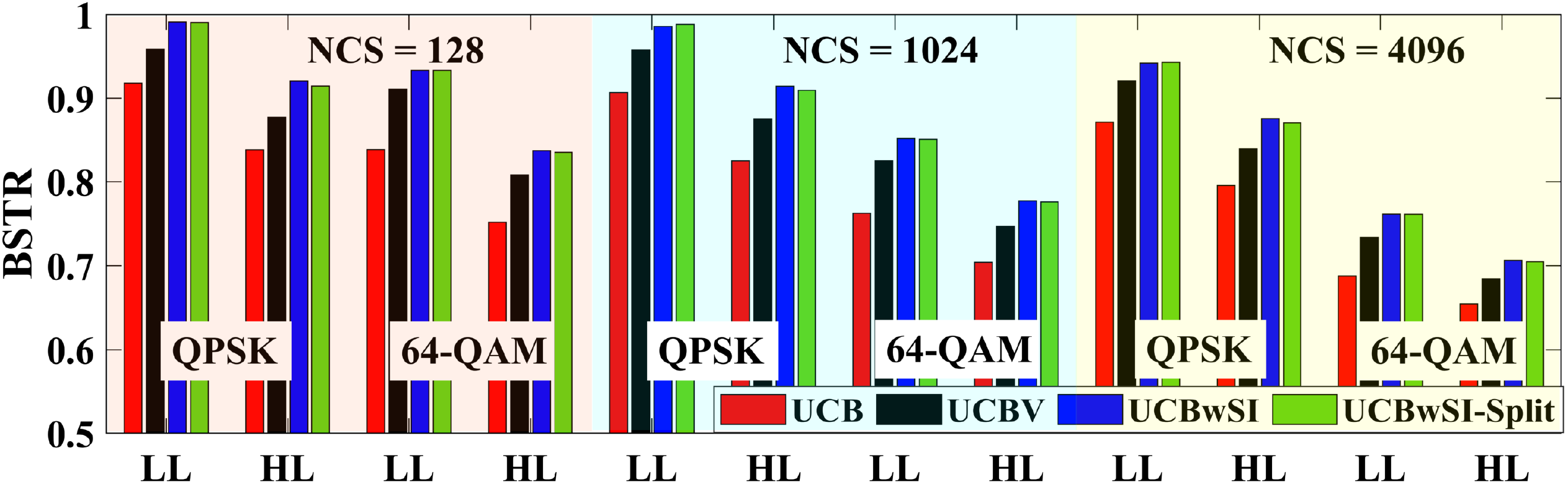}
    \caption{Effect of modulation scheme \& number of subcarriers on BSTR}
    \label{fig:OFDM_modfft_comp}
 
\end{figure}

\noindent
\textbf{Effect of modulation scheme and NCS on BSTR:} We consider two modulation schemes: 1) Quadrature phase shift keying (QPSK) which maps two bits in one data symbol, and 2) 64-Quadrature amplitude modulation (64-QAM) which maps six bits in one data symbol. In addition, we consider three NCS: $\{128, 1024, 4096\}$. In Fig.~\ref{fig:OFDM_modfft_comp}, it can be observed that the achieved BSTR is higher when the channel SNR variance is low. For a given NCS, the achieved BSTR using the QPSK is higher than the BSTR using the 64-QAM. This is expected since QPSK offers higher tolerance to channel distortions compared to 64-QAM. Similarly, with the increase in NCS from 128 to 4096, there is slight degradation in the BSTR. This is mainly due to fewer number of guard and pilot subcarriers which results in poor channel estimation and equalization at the receiver. This can be improved by increasing the number of guard and pilot subcarriers.

\begin{table*}[!h]
\caption{Comparison of Average Throughput in MBPS}
\centering
\label{tab:throughput_comp}
\resizebox{0.93\textwidth}{!}{%
\begin{tabular}{|c|c|c|c|c|c|c|c|c|c|c|c|c|c|c|c|c|}
\hline
\textbf{No. of Subcarriers} & \multicolumn{5}{c|}{\textbf{128}} & \multicolumn{5}{c|}{\textbf{1024}} & \multicolumn{5}{c|}{\textbf{4096}} & \multirow{3}{*}{\textbf{Average}} \\ \cline{1-16}
\textbf{Modulation Scheme} & \multicolumn{2}{c|}{\textbf{QPSK}} & \multicolumn{2}{c|}{\textbf{64-QAM}} & \multirow{2}{*}{\textbf{Average}} & \multicolumn{2}{c|}{\textbf{QPSK}} & \multicolumn{2}{c|}{\textbf{64-QAM}} & \multicolumn{1}{l|}{\multirow{2}{*}{\textbf{Average}}} & \multicolumn{2}{c|}{\textbf{QPSK}} & \multicolumn{2}{c|}{\textbf{64-QAM}} & \multicolumn{1}{l|}{\multirow{2}{*}{\textbf{Average}}} &  \\ \cline{1-5} \cline{7-10} \cline{12-15}
\textbf{SNR Variance} & \textbf{H} & \textbf{L} & \textbf{H} & \textbf{L} &  & \textbf{H} & \textbf{L} & \textbf{H} & \textbf{L} & \multicolumn{1}{l|}{} & \textbf{H} & \textbf{L} & \textbf{H} & \textbf{L} & \multicolumn{1}{l|}{} &  \\ \hline
\textbf{UCB} & 2.45 & 2.93 & 5.90 & 7.35 & 4.66 & 20.7 & 25 & 45.1 & 53 & 35.93 & 77.5 & 93 & 157.5 & 173 & 125.5 & 55.4 \\ \hline
\textbf{UCBV} & 2.68 & 3.20 & 6.82 & 8.66 & 5.34 & 23.3 & 27.9 & 50.8 & 62 & 41 & 86.4 & 103.8 & 172.1 & 198 & 140 & 62.1 \\ \hline
\textbf{UCB\_CV} & \textbf{2.95} & \textbf{3.42} & \textbf{7.33} & \textbf{9.10} & \textbf{5.70} & \textbf{25.4} & 29.5 & \textbf{55} & \textbf{66.1} & \textbf{44} & \textbf{94} & 108.7 & \textbf{183.5} & \textbf{213.1} & \textbf{149.8} & \textbf{66.5} \\ \hline
\textbf{Splitting} & 2.91 & \textbf{3.42} & 7.30 & \textbf{9.10} & 5.68 & 25.1 & \textbf{29.7} & 54.8 & 66 & 43.9 & 92.9 & \textbf{108.9} & 182.5 & 212.9 & 149.3 & 66.3 \\ \hline
\end{tabular}%
}
\vspace{-0.2cm}
\end{table*}
\noindent
\textbf{Effect of modulation scheme and NCS on Throughput:} We consider the achieved throughput which takes into account the BSTR along with the transmission time. For instance, higher BSTR in QPSK may not correspond to higher throughput. This is because, in a symbol time duration, QPSK transmits only two bits compared to six bits in 64-QAM.  The throughput results for different combinations of modulation scheme and NCS are given in Table~\ref{tab:throughput_comp}. Note that the throughput is calculated by taking into account the re-transmission time required so that all the message bits are correctly decoded at the receiver. As expected, throughput in case of channels with low SNR variance is higher than the channels with high SNR variance. Furthermore, throughput of 64-QAM is higher than QPSK and throughput increases significantly with the increase in the NCS. In all cases, UCBwSI offers significantly higher throughput than the UCB and UCBV algorithms. In an ideal noise-free distortion-less channel. throughput using the 64-QAM is almost 3-times the throughput using QPSK. However, due to realistic channel conditions and effect of interference, the improvement using 64-QAM is of the factor $1.2$ to $2.5$.  

\begin{figure}[!h]
\centering
    \includegraphics[scale=.33]{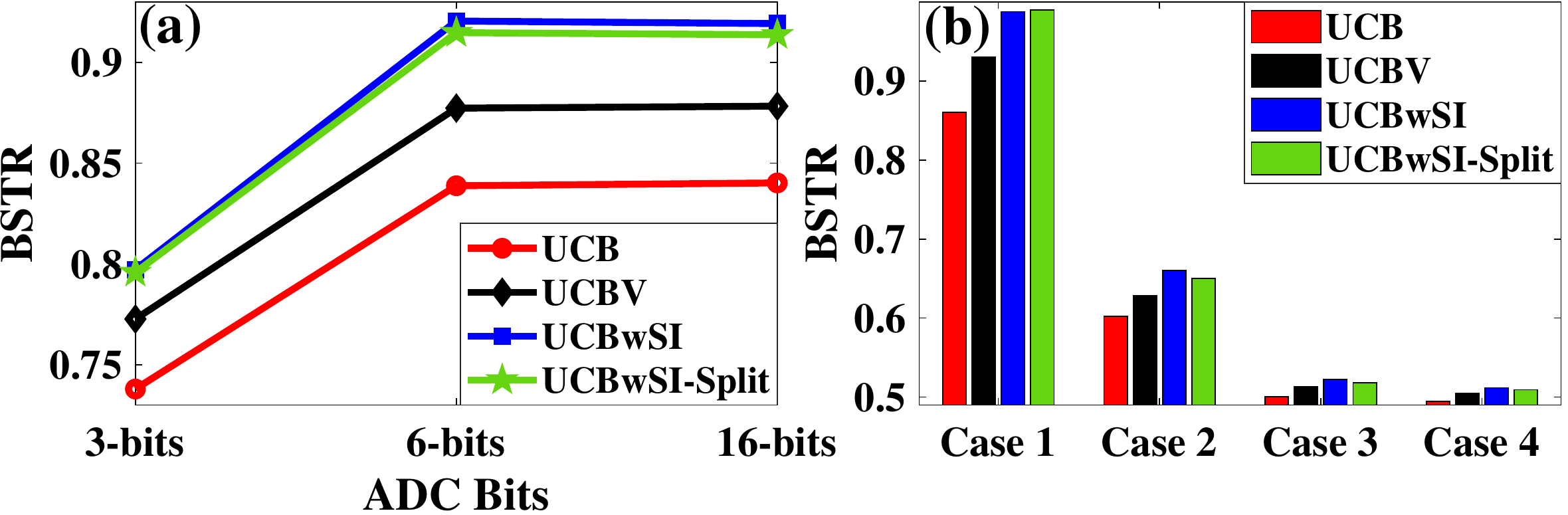}
    \vspace{-0.3cm}
    \caption{Effect of (a) ADC resolution and (b) Non-measurable interference on the achieved BSTR.}
    \label{fig:OFDM_ADCNMI_comp}
    \vspace{-0.3cm}
\end{figure}

\noindent
\textbf{Effect of ADC resolution on BSTR:}
At the receiver, an analog-to-digital converter (ADC) is used to digitize the signal received at the antenna. For ADC with $L$ bits of resolution, the maximum error due to quantization is $V/2^{L+1}$ where $V$ is the fixed full-scale voltage of ADC. To handle dynamically varying power levels, an automatic gain controller (AGC) is used before the ADC to scale the input signal so as to fit in the full-scale range of the ADC $[-V$, $V$]. In  Fig.~\ref{fig:OFDM_ADCNMI_comp} (a), we analyze the effect of $L$ on the achieved BSTR. For sufficiently large $L \geq 6$ and zero interference, there is no effect on the BSTR. When $L$ is small ($L=3$), the BSTR of all algorithms degrades significantly. This is because small $L$ needs substantial scaling by AGC which in turn degrades the received SNR significantly especially when high-power interference is present. The achieved BSTR depends on multiple factors out of which some are measurable with good accuracy. UCBwSI exploits such measurable factors and offers higher BSTR. 

\noindent
\textbf{Effect of Non-measurable SI on BSTR:}
Even though some of the SI are measurable, the achieved BSTR may not be significant compared to the UCB and UCBV algorithms if the unknown SI is dominant. To understand this behavior, we consider the out-of-band interference in the CR network. Specifically, we consider two interfering transmissions on the channels adjacent to the channel selected by CR user. One of the interference power levels is measurable at the transmitter while the other is not measurable or even unknown. We consider four cases for Non-measurable interference : 1) Case 1: weak 2) Case 2: slightly stronger than measurable interference but its carrier frequency is away from CR user 3) Case 3 and Case 4: much stronger than the measurable interference and its carrier frequency is close to CR user. As shown in Fig.~\ref{fig:OFDM_ADCNMI_comp} (b), achieved BSTR degrades with an increased impact of the non-measurable interference. Furthermore, the gap between the BSTR of the proposed algorithms and UCB and UCBV algorithms increases as the non-measurable interference becomes weaker than the measurable interference.

\begin{figure*}[!t]
\centering
    \includegraphics[width=0.9\linewidth]{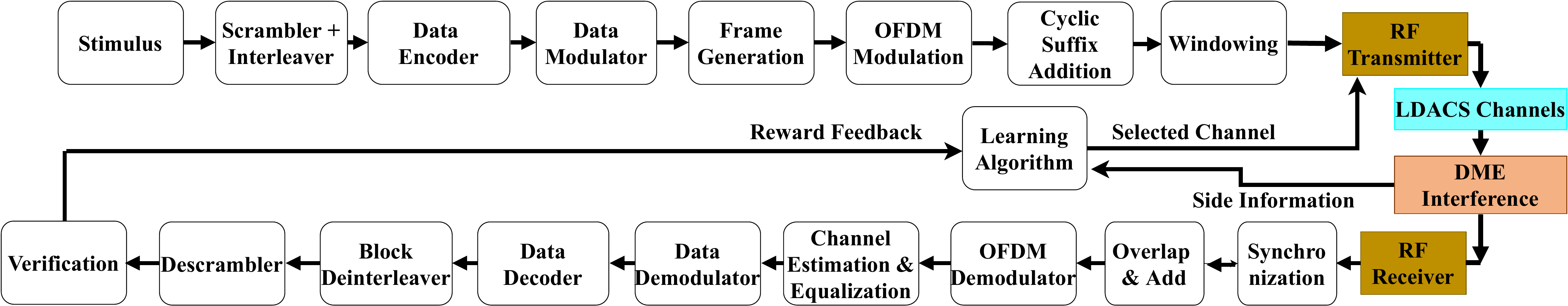}
    \caption{Block diagram showing different components of the LDACS physical layer at the transmitter and receiver.}
    \label{fig:LDACS_PHY}
\end{figure*}

\section{Application to LDACS}
\label{sec: LDACS}
We consider $L-$band  (960-1164 MHz) digital aeronautical communication system (LDACS) which is proposed as an alternative to the existing very high frequency (VHF) band (118-137 MHz) based air-to-ground communication system \cite{LDACS1,LDACS2}. It aims to utilize $1$ MHz vacant frequency bands between incumbent Distance Measuring Equipment (DME) signals in $L$-band. For instance, LDACS supports air-to-ground communications between aircraft and airport ground terminals and air-to-air communication between aircraft. Though the average transmit power levels of the DME signals are specified in the standards, the instantaneous power levels of the DME may vary dynamically due to moving aircraft \cite{LDACSIMI}.
In practical deployment, dynamically changing DME power levels significantly impact the BSTR and hence, the LDACS users need intelligence to identify a suitable $1$ MHz band to improve BSTR and throughput.  In this section, we demonstrate the application of UCBwSI to quickly identify the optimal frequency band using the DME power levels as side-information.

We realized the end-to-end LDACS physical layer as per the standards \cite{LDACS1} and corresponding building blocks are given in Fig.~\ref{fig:LDACS_PHY}. The LDACS physical layer parameters are summarized in Table~\ref{tab:LDACS_PHY}. The scrambled message bits are passed through a channel encoder consisting of sequential Reed-Solomon (RS) and convolution code (CC) based encoders with appropriate interleaving operations. This is followed by data modulation and waveform modulation using OFDM with windowing. At the receiver, the first step is timing and frequency synchronization via synchronization signals followed by interference mitigation. Then, wiener filtering-based channel estimation and zero-forcing-based channel equalization are performed using pilot signals. Thereafter, waveform demodulation, data demodulation, and channel decoding operations are performed. Detailed specifications can be found in \cite{LDACSStandard}.

\begin{table}[!h]
\caption{LDACS Physical Layer Parameters}
\vspace{-0.2cm}
\label{tab:LDACS_PHY}
\resizebox{0.47\textwidth}{!}{%
\begin{tabular}{|l|l|}
\hline
\textbf{Parameters} & \textbf{Value} \\ \hline
\textbf{Carrier Frequency} & 989.5 - 996.5 MHz \\ \hline
\textbf{Effective Bandwidth} & 498.05 KHz \\ \hline
\textbf{Sub-carrier Spacing} & 9.765625 KHz \\ \hline
\textbf{No. of Subcarriers} & 64 \\ \hline
\textbf{No. of Data Subcarriers} & 50 \\ \hline
\textbf{OFDM Symbols/Time slot} & 2 \\ \hline
\textbf{Time Slot Duration ($\mu$s)} & 120 \\ \hline
\textbf{Encoder Rate (CC, RS)} & \{0.5, 0.9\} \\ \hline
\textbf{Modulation Scheme} & \{QPSK,16-QAM,64-QAM\} \\ \hline
\textbf{(Bipolar) ADC Bits} & \{6, 16\} \\ \hline
\textbf{Wireless Channel} & \{ENT,APT,TMA\} \\ \hline
\textbf{Wireless Channel Parameters} & Refer to Table~\ref{tab:LDACS_WC} \\ \hline
\textbf{Out-of-band Interference} & DME \\ \hline
\textbf{DME Power Level} & As per the distribution \\ \hline
\end{tabular}%
}
\end{table}

\begin{figure}[!h]
\vspace{-0.3cm}
    \includegraphics[width=\linewidth]{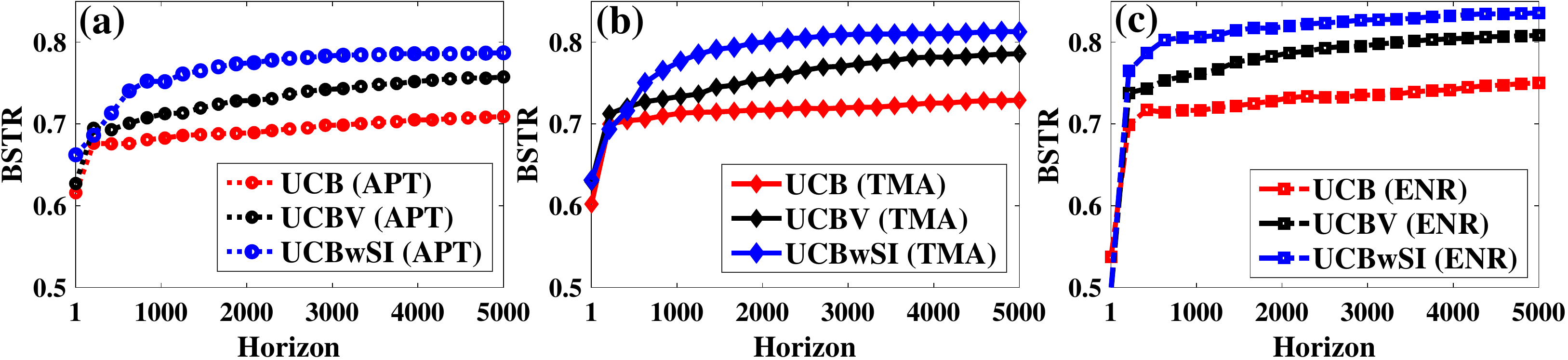}
    \vspace{-0.3cm}
    \caption{Effect of LDACS wireless channels on the achieved BSTR.}
    \label{fig:LDACS_WC}
    \vspace{-0.3cm}
\end{figure}

\noindent
\textbf{Effect of Wireless Channels on BSTR:}
In the LDACS environment, three types of wireless channels are experienced by users: 1) Airport (APT): During aircraft taxiing, 2) Terminal Maneuvering Area (TMA): During landing and take-off, and 3) En-routing (ENR): During flying phase. They are modeled as wide
sense stationary uncorrelated scattering channels and characterized as shown in Table~\ref{tab:LDACS_WC}. In Fig.~\ref{fig:LDACS_WC}, the performance of UCB, UCBV, and UCBwSI for three wireless channels is compared. Since the performance of UCBwSI and UCBwSI-Split is nearly identical, we have skipped UCBwSI-Split to avoid cluttering in plots. For all three channels, UCBwSI outperforms UCB and UCBV. Also, achieved BSTR is highest for ENR channel followed by TMA and APT. This is due to the presence of a strong line-of-sight (LoS), weak LoS and no LoS paths in the ENR, TMA and APT channels, respectively. 

\begin{table}[!h]
\vspace{-0.2cm}
\centering
\caption{LDACS Wireless Channels and Their Parameters}
\vspace{-0.2cm}
\label{tab:LDACS_WC}
\resizebox{0.5\textwidth}{!}{%
\begin{tabular}{|c|c|c|c|c|}
\hline
\textbf{Parameters} & \multirow{2}{*}{\textbf{\begin{tabular}[c]{@{}c@{}}Doppler \\ Frequency\end{tabular}}} & \multirow{2}{*}{\textbf{Fading}} & \multirow{2}{*}{\textbf{\begin{tabular}[c]{@{}c@{}}Rician \\ Factor\end{tabular}}} & \multirow{2}{*}{\textbf{Delay}} \\ \cline{1-1}
\multicolumn{1}{|l|}{\textbf{Channels}} &  &  &  &  \\ \hline
\textbf{ENR} & 1280 Hz & Rician & 15 dB & Direct + 2 delayed paths (0.3 $\mu$s, 15 $\mu$s) \\ \hline
\textbf{TMA} & 639 Hz & Rician & 10 dB & \begin{tabular}[c]{@{}c@{}}Exponentially decaying power delay profile\\ with a maximum delay of  20 $\mu$s\end{tabular} \\ \hline
\textbf{APT} & 423 Hz & Rayleigh & -100 dB & \begin{tabular}[c]{@{}c@{}}Exponentially decaying power delay profile\\ with a maximum delay of  3 $\mu$s\end{tabular} \\ \hline
\end{tabular}%
}
\end{table}

\begin{figure}[!h]
\vspace{-0.2cm}
    \includegraphics[width=\linewidth]{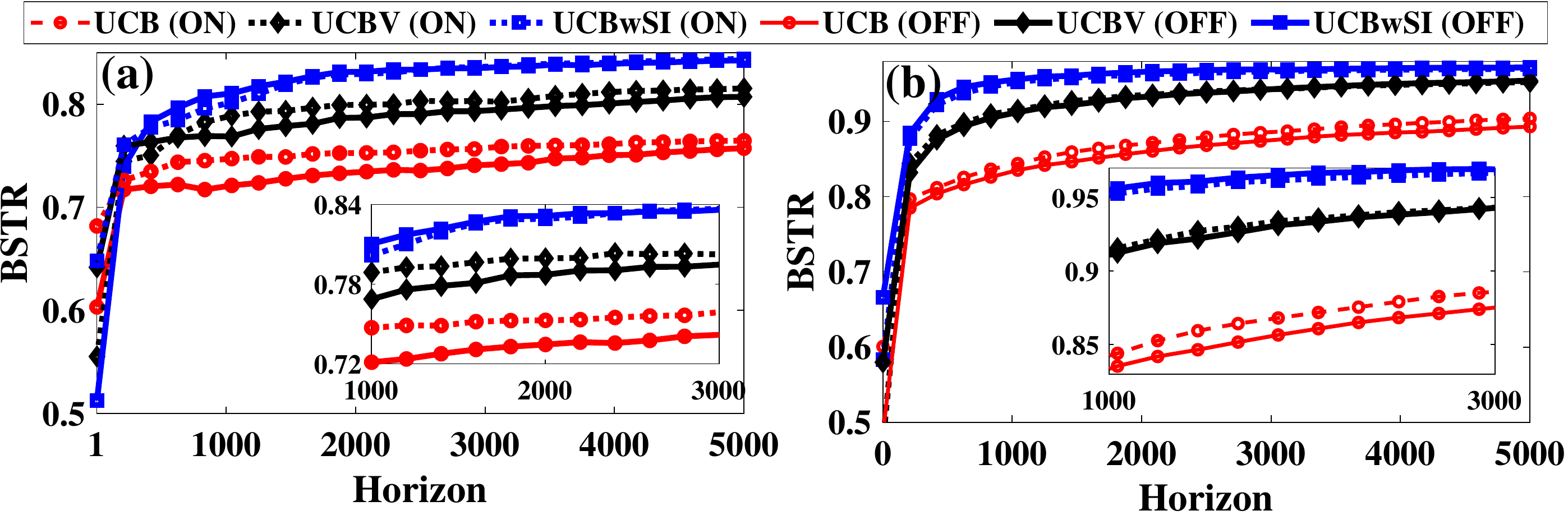}
    \caption{Effect of DME interference mitigation in LDACS on achieved BSTR.}
    \label{fig:LDACS_IM}
    \vspace{-0.3cm}
\end{figure}

\noindent
\textbf{Effect of Interference Mitigation on BSTR:}
Various simulation results have shown slight degradation in the BSTR when the DME signal power level in the adjacent channel is high. To mitigate the interference, the pulse banking approach has been incorporated recently in the LDACS standard. In Fig.~\ref{fig:LDACS_IM}, we analyzed the effect of DME interference mitigation techniques in existing LDACS on the achieved BSTR. Fig.~\ref{fig:LDACS_IM} (a) and (b) consider the LDACS ENR channel with high and low SNR variance, respectively. As expected, the achieved BSTR is higher when variance is low.  Furthermore, UCBwSI offers significantly higher BSTR than UCB and UCBV. The improvement after enabling the interference mitigation at the receiver is significant for UCB and UCBV compared to UCBwSI as UCBwSI already exploits the DME SI to select channels. Thus, our approach can eliminate additional interference mitigation at the receiver which can potentially improve the area, power, and latency of the LDACS physical layer. 

\begin{figure}[!h]
    \includegraphics[width=0.95\linewidth]{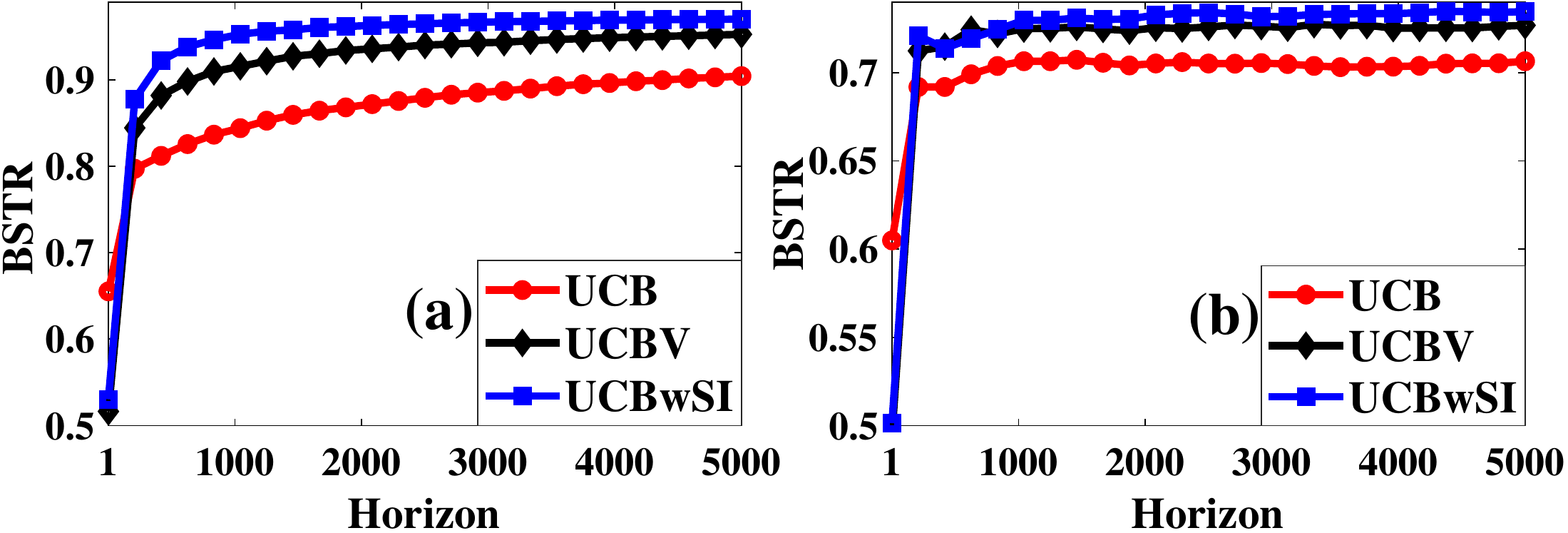}
    \vspace{-0.2cm}
    \caption{Effect of number of arms, (a) $K=3$, and (b) $K=8$ on BSTR.}
    \label{fig:LDACS_arms}
\end{figure}

\noindent
\textbf{Effect of number of channels on BSTR:} With the increase in the number of available channels, exploration time increases, and hence, learning algorithms need more time to identify the optimal channel. In Fig.~\ref{fig:LDACS_arms} (a)  and (b), the BSTR is compared for $K=3$ and $K=8$, respectively. As expected, exploration time is significantly lower for $K=3$. Furthermore, the difference between the performance UCBwSI and UCB/UCBV increases with $K$. This is because the proposed UCBwSI algorithm exploits SI to reduce exploration time.

\section{Conclusions}
\label{sec: Conclusion}

We developed improved learning algorithms that exploited side-information in wireless networks. We showed that side-information helps to compute better point-estimator that have smaller variance and confidence intervals when their mean values are known. We developed two UCB based algorithms named UCBwSI and UCBwSI-Split that incorporate side-information and achieve better regret performance. The UCBwSI is designed under the assumption that reward and side-information are jointly Gaussian and UCBwSI-Split works for arbitrary distributions and is based on a re-sampling approach. The improvement in regret depends on the correlation between the reward and side-information of each arm, and the higher the correlation smaller is the regret. 

We showed through extensive and realistic simulation of cognitive radio networks and $L$-band digital aeronautical communication system that side-information can be exploited to improve network throughput. Specifically, our experiments demonstrated that the correlation between side-information and reward is often strong in networks and results in significant performance gains. It is interesting to apply our setup on the 5G network and quantify the inefficiency when the mean of side-information is known approximately or has to be estimated.

\section*{Acknowledgements}
Manjesh K. Hanawal is supported by INSPIRE faculty fellowship from
DST and Early Career Research Award (ECR/2018/002953) from SERB,
Govt. of India. S

\appendices

\section{Control Variates}
\label{app: ControlVariates}

Let $\mu$ be the unknown quantity that needs to be estimated, and $X$ be an unbiased estimator of $\mu$, i.e., $\EE{X} = \mu$. A random variable $W$ is a CV if its expectation $\omega$ is known and it is correlated with $X$. Linear control methods use errors in estimates of known random variables to reduce error in estimation of unknown random variable. For any choice of a coefficient $\beta$, define a new  estimator as
$
    \bar{X} = X + \beta(\omega-W).
$
It is straightforward to verify that its variance is given by
\al{
    \text{Var}(\bar{X}) 
    &= \text{Var}(X) + \beta^2\text{Var}(W) - 2\beta\text{Cov}(X,W).  \label{equ:varNewEst}
}
and is minimized by setting $\beta$ to $\beta^\star = \text{Cov}(X,W)/\text{Var}(W)$.
The minimum value of the variance is given by $\text{Var}(\bar{X}) = (1 - \rho^2)\text{Var}(X)$,
where $\rho$ is the correlation coefficient of $X$ and $W$. Larger the correlation, the greater the variance reduction achieved by the CV. In practice, the values of $\text{Cov}(X,W)$ and $\text{Var}(W)$ are unknown and need to be estimated to compute the best approximation for $\beta^\star$.

\section{Incorporating Multiple Side-information}
\label{app: MultipleCV}
In some applications it could be possible more than one side-information could be availble. We denote the number of side-information with each arm as $q$. Let $W_{t,i,j}$ be the $j^{\text{th}}$ side-information of arm $i$ that is observed in round $t$, . Then the unbiased mean reward estimator for arm $i$ with associated side-information is given by $\hat\mu_{s,i,q}^c = \hat\mu_{s,i} + \boldsymbol{\hat\beta}_{i}^{*\top} (\boldsymbol{\omega}_{i} - \boldsymbol{\hat\omega}_{s,i})$
where $\boldsymbol{\hat\beta}^*_i = \left(\hat\beta_{i,1}, \ldots, \hat\beta_{i,q}\right)^\top$, $\boldsymbol{\omega}_{i} = \left(\omega_{i,1}, \ldots, \omega_{i,q}\right)^\top$, and $\boldsymbol{\hat\omega}_{s,i}= \left(\hat\omega_{s,i,1}, \ldots, \hat\omega_{s,i,q}\right)^\top$.
Let $s$ be the number of rewards and associated side-information  samples for arm $i$,  $\boldsymbol{W}_i$ be the $s\times q$ matrix whose $r^{\text{th}}$ row is $\left(W_{r,i,1}, W_{r,i,2}, \ldots, W_{r,i,q} \right)$, and $\boldsymbol{X}_i=(X_{1,i}, \ldots, X_{s,i})^\top$. By extending the arguments used in \cref{equ:estBeta} to $q$ side-information, the estimated coefficient vector is given by $\boldsymbol{\hat\beta}^*_i = (\boldsymbol{W}_i^\top\boldsymbol{W}_i - s\boldsymbol{\hat\omega}_{s,i}\boldsymbol{\hat\omega}_{s,i}^\top )^{-1} (\boldsymbol{W}_i^\top \boldsymbol{X}_i - s{\boldsymbol{\hat\omega}_i}~\hat\mu_{s,i}).$
We can generalize this for MAB problems with $q$ side-informations and then use \ref{alg:UCBwSI} with $Q=q+2$ and appropriate optimistic upper bound for multiple side-informations. 

%

\section{Proof of \cref{thm:regretBound}}
\label{app: ProofRegretBound}
\noindent
Following the standard UCB arguments, if $i\neq i^*$ is selected in round $t$,
at least of the event happens.
	\als{
		&E1.~ \hat\mu_{N_i^\star(t),i^\star}^c + V_{t,N_i^\star(t)}^{(\alpha)}\sqrt{{\hat\nu_{N_i^\star(t),i^\star}}}  \le \mu_{i^\star},\\
		&E2.~ \hat\mu_{N_i(t),i}^c - V_{t,N_i(t)}^{(\alpha)}\sqrt{{\hat\nu_{N_i(t),i}}} > \mu_i, \text{ and}\\
		&E3.~ N_i(t) < \frac{4(V_{T,N_i(t)}^{(\alpha)})^2 \hat\nu_{N_i(T),i}  N_i(T)}{\Delta_i^2}.
	}
Define $u=\ceil{\frac{4(V_{T,N_i(T)}^{(2)})^2  \hat\nu_{N_i(T),i}  N_i(T)}{\Delta_i^2}}$. 
Write $N_i(T)=\sum_{t=1}^T \mathds{1}_{\{I_t=i\}}$. we have
\[
		\mathbb{E}[{N_i(T)}]  \leq 	\mathbb{E}[u] + 	\mathbb{E}\left [{\sum_{t=1}^T \mathds{1}_{\{\text{E1 is true}\}}}\right] +  	\mathbb{E}\left [{\sum_{t=1}^T \mathds{1}_{\{\text{E2 is true}\}}}\right].  \label{equ:expectedPull} 
\]

	\al{
	&\mathbb{E}\left [{\sum_{t=1}^T \mathds{1}_{\{\text{E1 is true}\}}}\right] = \sum_{t=1}^T \Prob{\text{E1 is true}} \nonumber \\
	&= \sum_{t=1}^T \Prob{\hat\mu_{N_i^\star(t),i^\star,q}^c - \mu_{i^\star} \le - V_{t,N_i(t),q}^{(2)}\sqrt{{\hat\nu_{N_i^\star(t),i^\star,q}}}} \nonumber \\
	&\leq \sum_{t=1}^T \frac{1}{t^2} \le \sum_{t=1}^\infty \frac{1}{t^2}   \le \frac{\pi^2}{6},  \nonumber  \label{equ:expE1bound}
}
where the first inequality follows from \cref{prop:confProb} by setting $\alpha=2$. Following similar arguments we get $\mathbb{E}\left [{\sum_{t=1}^T \mathds{1}_{\{\text{E2 is true}\}}}\right]\leq \pi^2/6$.   Notice that $u$ is random as its depends on $V_{T,N_i(T),q}^{(2)}$ and $\hat\nu_{N_i(T),i}$. We bound $\mathbb{E}[u]$ as follows:
\als{
	&\mathbb{E}[u]  \leq \frac{4}{\Delta_i^2}\mathbb{E}\left[{(V_{T,N_i(T)}^{(2)})^2 \hat\nu_{N_i(T),i}  N_i(T)}\right] + 1.\\
	&= \frac{4(V_{T,T}^{(2)})^2}{\Delta_i^2}\mathbb{E}\left [{\left(\frac{V_{T,N_i(T)}^{(2)}}{V_{T,T}^{(2)}}\right)^2  \hat\nu_{N_i(T),i}  N_i(T)}\right] + 1\\
	&= \frac{4(V_{T,T}^{(2)})^2}{\Delta_i^2}\mathbb{E}\left [{\left(\frac{V_{T,N_i(T)}^{(2)}}{V_{T,T}^{(2)}}\right)^2  \mathbb{E}[\hat\nu_{N_i(T),i}]  N_i(T)}| N_i(T)\right] + 1 \\
	&= \frac{4(V_{T,T}^{(2)})^2}{\Delta_i^2}\mathbb{E}\left [{\left(\frac{V_{T,N_i(T)}^{(2)}}{V_{T,T}^{(2)}}\right)^2\hspace{-.3cm}  \frac{N_i(T)-2}{N_i(T)-3}(1-\rho_i^2)  \sigma_i^2}| N_i(T)\right] + 1 \\
	& \quad (\text{From \cref{prop:varEst} and \cref{prop:estProp}})\\
		&\leq \frac{8(V_{T,T}^{(2)})^2}{\Delta_i^2}(1-\rho_i^2)  \sigma_i^2\mathbb{E}\left [{\left(\frac{V_{T,N_i(T)}^{(2)}}{V_{T,T}^{(2)}}\right)^2}\right] + 1 ~~\mbox{as} ~~N_i(T)\geq 4.
	} 
The final regret bound follows from plugging the above in $\Regret_T=\sum_{i}\Delta_i\mathbb{E}[N_i(T)]$.	

\newpage
\bibliography{ref}

\begin{thebibliography}{10}
\providecommand{\url}[1]{#1}
\csname url@samestyle\endcsname
\providecommand{\newblock}{\relax}
\providecommand{\bibinfo}[2]{#2}
\providecommand{\BIBentrySTDinterwordspacing}{\spaceskip=0pt\relax}
\providecommand{\BIBentryALTinterwordstretchfactor}{4}
\providecommand{\BIBentryALTinterwordspacing}{\spaceskip=\fontdimen2\font plus
\BIBentryALTinterwordstretchfactor\fontdimen3\font minus
  \fontdimen4\font\relax}
\providecommand{\BIBforeignlanguage}[2]{{%
\expandafter\ifx\csname l@#1\endcsname\relax
\typeout{** WARNING: IEEEtran.bst: No hyphenation pattern has been}%
\typeout{** loaded for the language `#1'. Using the pattern for}%
\typeout{** the default language instead.}%
\else
\language=\csname l@#1\endcsname
\fi
#2}}
\providecommand{\BIBdecl}{\relax}
\BIBdecl

\bibitem{JSAC2017_MillimeterWaves}
M.~Xiao, S.~Mumtaz, Y.~Huang, L.~Dai, Y.~Li, M.~Matthaiou, G.~K. Karagiannidis,
  E.~Björnson, K.~Yang, C.-L. I, and A.~Ghosh, ``Millimeter wave
  communications for future mobile networks,'' \emph{IEEE Journal on Selected
  Areas in Communications (JSAC)}, vol.~35, no.~9, pp. 1909--1935, 2017.

\bibitem{JSTSP2011_CognitiveRadioSurvey_BeibeiRay}
B.~Wang and K.~R. Liu, ``Advances in cognitive radio networks: A survey,''
  \emph{IEEE Journal of Selected Topics in Signal Processing}, vol.~5, no.~1,
  pp. 5--23, 2011.

\bibitem{JNCA2014_AdHocNetworkSurvey_AlSultan}
S.~Al-Sultan, M.~M. Al-Doori, A.~H. Al-Bayatti, and H.~Zedan, ``A comprehensive
  survey on vehicular ad hoc network,'' \emph{Journal of Network and Computer
  Applications}, vol.~37, pp. 380--392, 2014.

\bibitem{NRbulletsbook}
C.~Johnson, in \emph{5G New Radio in Bullets}.\hskip 1em plus 0.5em minus
  0.4em\relax [Online] https://books.google.co.in/books?id=NoRjyAEACAAJ, 2019.

\bibitem{5G_interference_measurement}
H.~Elgendi, M.~Mäenpää, T.~Levanen, T.~Ihalainen, S.~Nielsen, and
  M.~Valkama, ``Interference measurement methods in 5g nr: Principles and
  performance,'' in \emph{2019 16th International Symposium on Wireless
  Communication Systems (ISWCS)}, 2019, pp. 233--238.

\bibitem{JSAC11_DistributedLearning_Anadakumar}
A.~Anandkumar, N.~Michael, A.~K. Tang, and A.~Swami, ``Distributed algorithms
  for learning and cognitive medium access with logarithmic regret,''
  \emph{IEEE Journal on Selected Areas in Communications}, vol.~29, no.~4, pp.
  731--745, 2011.

\bibitem{TIT14_DecentralizedLearning_KalthilNayyarJain}
D.~Kalathil, N.~Nayyar, and R.~Jain, ``Decentralized learning for multiplayer
  multiarmed bandits,'' \emph{IEEE Transactions on Information Theory},
  vol.~60, no.~4, pp. 2331--2345, 2014.

\bibitem{TCNS2018_DecentralizedLearning_KalthilNayyarJain}
N.~Nayyar, D.~Kalathil, and R.~Jain, ``On regret-optimal learning in
  decentralized multiplayer multiarmed bandits,'' \emph{IEEE Transactions on
  Control of Network Systems}, vol.~5, no.~1, pp. 597--606, 2018.

\bibitem{Infocom2019_DistributedLearning_TibrewalPatchalaHanawal}
H.~Tibrewal, S.~Patchala, M.~K. Hanawal, and S.~J. Darak, ``Distributed
  learning and optimal assignment in multiplayer heterogeneous networks,'' in
  \emph{IEEE International Conference on Computer Communications (INFOCOM)},
  2019.

\bibitem{MCL2021_MABEnergyHavesting}
D.~Ghosh, M.~K. Hanawal, and N.~Zlatanov, ``Learning to optimize energy
  efficiency in energy harvesting wireless sensor networks,'' \emph{IEEE
  Wireless Communications Letters}, vol.~10, no.~6, pp. 1153--1157, 2021.

\bibitem{JSAC2015_MABEnergyHarvesting}
P.~Blasco and D.~Gündüz, ``Multi-access communications with energy
  harvesting: A multi-armed bandit model and the optimality of the myopic
  policy,'' \emph{IEEE Journal on Selected Areas in Communications}, vol.~33,
  no.~3, pp. 585--597, 2015.

\bibitem{TIT2018_OnlineLearningEnegyHarvesting}
P.~Sakulkar and B.~Krishnamachari, ``Online learning schemes for power
  allocation in energy harvesting communications,'' \emph{IEEE Transactions on
  Information Theory}, vol.~64, no.~6, pp. 4610--4628, 2018.

\bibitem{INFOCOM2020_MAMBA}
I.~Aykin, B.~Akgun, M.~Feng, and M.~Krunz, ``Mamba: A multi-armed bandit
  framework for beam tracking in millimeter-wave systems,'' in \emph{IEEE
  Conference on Computer Communications (INFOCOM)}, 2020, pp. 1469--1478.

\bibitem{INFOCOM2020_OnlineBayesianLearningmmWave}
M.~A. Qureshi and C.~Tekin, ``Online bayesian learning for rate selection in
  millimeter wave cognitive radio networks,'' in \emph{IEEE Conference on
  Computer Communications (INFOCOM)}, 2020, pp. 1449--1458.

\bibitem{VTC2018_MAB5G}
A.~Vora and K.-D. Kang, ``Throughput enhancement via multi-armed bandit in
  heterogeneous 5g networks,'' in \emph{2018 IEEE 88th Vehicular Technology
  Conference (VTC-Fall)}, 2018, pp. 1--5.

\bibitem{TCOM2021_MABNOMA}
M.-J. Youssef, V.~V. Veeravalli, J.~Farah, C.~A. Nour, and C.~Douillard,
  ``Resource allocation in noma-based self-organizing networks using stochastic
  multi-armed bandits,'' \emph{IEEE Transactions on Communications}, 2021.

\bibitem{SSPW2021_MABNOMA}
A.~H.~E. Hassani, A.~Savard, and E.~V. Belmega, ``Energy-efficient 1-bit
  feedback noma in wireless networks with no csit/cdit,'' in \emph{2019 IEEE
  Wireless Communications and Networking Conference Workshop (WCNCW)}, 2021.

\bibitem{ML02_auer2002finite}
P.~Auer, N.~Cesa-Bianchi, and P.~Fischer, ``Finite-time analysis of the
  multiarmed bandit problem,'' \emph{Machine learning}, pp. 235--256, 2002.

\bibitem{COLT11_garivier2011kl}
A.~Garivier and O.~Capp{\'e}, ``The kl-ucb algorithm for bounded stochastic
  bandits and beyond,'' in \emph{Proceedings of the 24th annual Conference On
  Learning Theory}, 2011, pp. 359--376.

\bibitem{AISTATS2013_RegretBoundsForThomsonSampling_AgarwalGoyal}
S.~Agrawal and N.~Goyal, ``Further optimal regret bounds for thompson
  sampling,'' in \emph{16th International Conference on Artificial Intelligence
  and Statistics (AISTATS)}, Scottsdale, USA, 2013.

\bibitem{Book_OnlineLearningNetworks}
C.~Tekin and M.~Liu, \emph{Online Learning Methods for Networking}.\hskip 1em
  plus 0.5em minus 0.4em\relax {NOW} publisher, Foundations and Trends in
  Networking, 2015.

\bibitem{WirelessCommunications2020_MAB_Survey}
F.~Li, D.~Yu, H.~Yang, J.~Yu, H.~Karl, and X.~Cheng, ``Multi-armed-bandit-based
  spectrum scheduling algorithms in wireless networks: A survey,'' \emph{IEEE
  Wireless Communications}, vol.~27, no.~1, pp. 24--30, 2020.

\bibitem{WirelessCommunication2016_MAB5G}
S.~Maghsudi and E.~Hossain, ``Multi-armed bandits with application to {5G}
  small cells,'' \emph{IEEE Wireless Communications}, vol.~23, no.~3, pp.
  64--73, 2016.

\bibitem{INFOCOM2014_OptimalRateSampling_Combes}
R.~Combes, A.~Proutiere, D.~Yun, J.~Ok, and Y.~Yi, ``Optimal rate sampling in
  802.11 systems,'' in \emph{IEEE INFOCOM 2014 - IEEE Conference on Computer
  Communications}, 2014, pp. 2760--2767.

\bibitem{TMC2019_OptimalRateSampling_Combes}
R.~Combes, J.~Ok, A.~Proutiere, D.~Yun, and Y.~Yi, ``Optimal rate sampling in
  802.11 systems: Theory, design, and implementation,'' \emph{IEEE Transactions
  on Mobile Computing}, vol.~18, no.~5, pp. 1145--1158, 2019.

\bibitem{INFOCOM2018_MABmmWave_HashemiSbharwalShroff}
M.~Hashemi, A.~Sabharwal, C.~Emre~Koksal, and N.~B. Shroff, ``Efficient beam
  alignment in millimeter wave systems using contextual bandits,'' in
  \emph{IEEE Conference on Computer Communications (INFOCOM)}, 2018, pp.
  2393--2401.

\bibitem{TWC2019_FastmmWave}
W.~Wu, N.~Cheng, N.~Zhang, P.~Yang, W.~Zhuang, and X.~Shen, ``Fast mmwave beam
  alignment via correlated bandit learning,'' \emph{IEEE Transactions on
  Wireless Communications}, vol.~18, no.~12, pp. 5894--5908, 2019.

\bibitem{MACS2018_MABSideObservations_YunProutiere}
D.~Yun, A.~Proutiere, S.~Ahn, J.~Shin, and Y.~Yi, ``Multi-armed bandit with
  additional observations,'' \emph{Measurement and Analysis of Computing
  Systems}, vol.~2, no.~1, Apr. 2018.

\bibitem{ITW2013_MABSpectrumBandit_LeLargeProtiere}
M.~Lelarge, A.~Proutiere, and M.~S. Talebi, ``Spectrum bandit optimization,''
  in \emph{IEEE Information Theory Workshop (ITW)}, 2013, pp. 1--5.

\bibitem{NetworkMeets_ContextualMAB}
V.~Saxena, J.~Jald\'{e}n, J.~E. Gonzalez, M.~Bengtsson, H.~Tullberg, and
  I.~Stoica, ``Contextual multi-armed bandits for link adaptation in cellular
  networks,'' in \emph{Proceedings of the Workshop on Network Meets AI \& ML},
  2019, p. 44–49.

\bibitem{ICOIN2021_ContextualBanditIndustry4.0}
R.~Bajrachrya and H.~Jung, ``Contextual bandits approach for selecting the best
  channel in industry 4.0 network,'' in \emph{2021 International Conference on
  Information Networking (ICOIN)}, 2021, pp. 13--16.

\bibitem{JORS85_james1985variance}
B.~James, ``Variance reduction techniques,'' \emph{Journal of the Operational
  Research Society}, vol.~36, no.~6, pp. 525--530, 1985.

\bibitem{OR90_nelson1990control}
B.~L. Nelson, ``Control variate remedies,'' \emph{Operations Research},
  vol.~38, no.~6, pp. 974--992, 1990.

\bibitem{MS82_lavenberg1981perspective}
S.~S. Lavenberg and P.~D. Welch, ``A perspective on the use of control
  variables to increase the efficiency of monte carlo simulations,''
  \emph{Management Science}, vol.~27, no.~3, pp. 322--335, 1981.

\bibitem{OR82_lavenberg1982statistical}
S.~S. Lavenberg, T.~L. Moeller, and P.~D. Welch, ``Statistical results on
  control variables with application to queueing network simulation,''
  \emph{Operations Research}, vol.~30, no.~1, pp. 182--202, 1982.

\bibitem{EJOR89_nelson1989batch}
B.~L. Nelson, ``Batch size effects on the efficiency of control variates in
  simulation,'' \emph{European Journal of Operational Research}, vol.~43,
  no.~2, pp. 184--196, 1989.

\bibitem{TCS09_audibert2009exploration}
J.-Y. Audibert, R.~Munos, and C.~Szepesv{\'a}ri, ``Exploration--exploitation
  tradeoff using variance estimates in multi-armed bandits,'' \emph{Theoretical
  Computer Science}, vol. 410, no.~19, pp. 1876--1902, 2009.

\bibitem{TCS09_UCBV_Audibert}
J.-Y. Audibert, R.~Munos, and C.~Szepesv\'ari, ``Exploration–exploitation
  tradeoff using variance estimates in multi-armed bandits,'' \emph{Theoretical
  Computer Science}, vol. 410, no.~19, pp. 1876 -- 1902, 2009.

\bibitem{IIE2001_BiasedControlVariate_Bruce}
B.~W. Schmeiser, M.~R. Taaffe, and J.~Wang, ``Biased control-variate
  estimation,'' \emph{IIE Transactions}, vol.~33, no.~3, pp. 219--228, 2001.

\bibitem{IIE2012_ControlVariate_Raghu}
R.~Pasupathy, B.~W. Schmeiser, M.~R. Taaffe, and J.~Wang, ``Control-variate
  estimation using estimated control means,'' \emph{IIE Transactions}, vol.~44,
  no.~5, pp. 381--385, 2012.

\bibitem{NRL91_avramidis1991simulation}
A.~N. Avramidis, K.~W. Bauer~Jr, and J.~R. Wilson, ``Simulation of stochastic
  activity networks using path control variates,'' \emph{Naval Research
  Logistics (NRL)}, vol.~38, no.~2, pp. 183--201, 1991.

\bibitem{LDACS1}
M.~Schnell, U.~Epple, D.~Shutin, and N.~Schneckenburger, ``Ldacs: future
  aeronautical communications for air-traffic management,'' \emph{IEEE
  Communications Magazine}, vol.~52, no.~5, pp. 104--110, 2014.

\bibitem{LDACS2}
C.~Rihacek, B.~Haindl, P.~Fantappie, S.~Pierattelli, T.~Gräupl, M.~Schnell,
  and N.~Fistas, ``L-band digital aeronautical communications system (ldacs)
  activities in sesar2020,'' in \emph{2018 Integrated Communications,
  Navigation, Surveillance Conference (ICNS)}, 2018, pp. 4A1--1--4A1--8.

\bibitem{LDACSIMI}
U.~Epple and M.~Schnell, ``Overview of interference situation and mitigation
  techniques for ldacs1,'' in \emph{2011 IEEE/AIAA 30th Digital Avionics
  Systems Conference}, 2011, pp. 4C5--1--4C5--12.

\bibitem{LDACSStandard}
U.~E. Miodrag~Sajatovic, Bernhard~Haindl and T.~Gräupl, ``Updated {LDACS1}
  system specification,''
  http://www.ldacs.com/wp-content/uploads/2014/02/LDACS1-Updated-Specification-Proposal-D2-Deliverable.pdf,
  2011.

\end{thebibliography}
\bibliographystyle{IEEEtran}
\end{document}